\documentclass[aps,prl,twocolumn,showpacs,superscriptaddress]{revtex4}  
\usepackage[]{graphicx}  
\usepackage{dcolumn}   
\usepackage{bm}        
\usepackage{amssymb}   

\usepackage{amsmath}
\usepackage{xspace}
\usepackage[thinspace,thinspace,squaren,textstyle]{SIunits}
\newcommand{\ie}{\mbox{i.\,e.}\xspace}
\newcommand{\lhs}{\mbox{l.\,h.\,s.}\xspace}
\newcommand{\rhs}{\mbox{r.\,h.\,s.}\xspace}

\begin{document}

\title{The critical velocity in the BEC-BCS crossover}

\begin{abstract}

We map out the critical velocity in the crossover from Bose-Einstein condensation (BEC) to Bardeen-Cooper-Schrieffer superfluidity with ultracold $^{6}$Li gases.
A small attractive potential is dragged along lines of constant column density. The rate of the induced heating increases steeply above a critical velocity $v_c$.
In the same samples, we measure the speed of sound $v_s$ by exciting density waves and compare the results to the measured values of $v_c$. We perform numerical simulations in the BEC regime and find very good agreement, validating the approach. In the strongly correlated regime, where theoretical predictions only exist for the speed of sound, our measurements of $v_c$ provide a testing ground for theoretical approaches.
\end{abstract}

\author{Wolf Weimer}
\author{Kai Morgener}
\affiliation{Institut f\"ur Laserphysik, Universit\"at Hamburg, Luruper Chaussee 149, 22761 Hamburg, Germany}
\author{Vijay Pal Singh}
\affiliation{Institut f\"ur Laserphysik, Universit\"at Hamburg, Luruper Chaussee 149, 22761 Hamburg, Germany}
\affiliation{Zentrum f\"ur Optische Quantentechnologien, Universit\"at Hamburg, Luruper Chaussee 149, 22761 Hamburg, Germany}
\affiliation{The Hamburg Centre for Ultrafast Imaging, Luruper Chaussee 149, 22761 Hamburg, Germany}
\author{Jonas Siegl}
\author{Klaus Hueck}
\author{Niclas Luick}
\affiliation{Institut f\"ur Laserphysik, Universit\"at Hamburg, Luruper Chaussee 149, 22761 Hamburg, Germany}
\author{Ludwig Mathey}
\affiliation{Institut f\"ur Laserphysik, Universit\"at Hamburg, Luruper Chaussee 149, 22761 Hamburg, Germany}
\affiliation{Zentrum f\"ur Optische Quantentechnologien, Universit\"at Hamburg, Luruper Chaussee 149, 22761 Hamburg, Germany}
\affiliation{The Hamburg Centre for Ultrafast Imaging, Luruper Chaussee 149, 22761 Hamburg, Germany}
\author{Henning Moritz}
\affiliation{Institut f\"ur Laserphysik, Universit\"at Hamburg, Luruper Chaussee 149, 22761 Hamburg, Germany}


\maketitle
\vskip 0.25cm 

\date{\today}


Frictionless flow of charged or neutral particles is one of the most striking macroscopic phenomena arising from quantum physics. Its appearance is remarkably widespread, ranging from superconductivity in solids to superfluidity in liquids and dilute gases with flow of either bosonic or fermionic particles. For technological applications, stability against thermal fluctuations or external perturbations is crucial. The corresponding quantities, \ie critical temperature and critical velocity, are typically highest in the strongly correlated regime, where the interactions stabilizing the many-body state are
particularly strong. Attaining a full understanding of the underlying
microscopic mechanisms in this regime is one of the major challenges
of modern physics. Ultracold atomic gases have emerged as an excellent
platform to study the influence of microscopic physics on macroscopic
observables \cite{Zwerger2008,Ketterle1999,further_SF,Dalibard2012}. 

\begin{figure}
\includegraphics[]{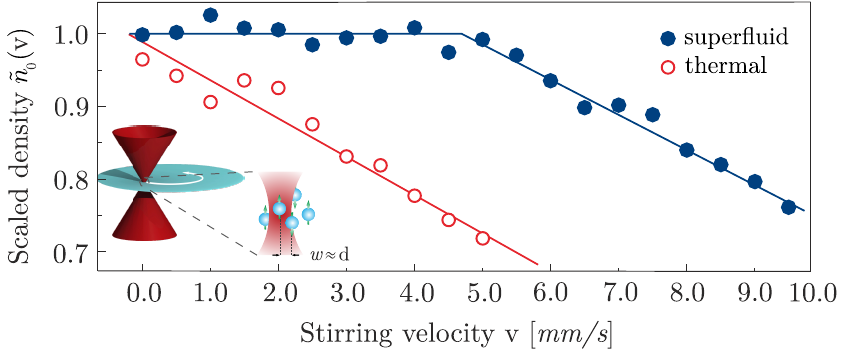}
\caption{\label{fig:scheme}
A red detuned laser beam with waist $w$ moves through the cloud with velocity $v$, where the obstacle size is on the order of the inter-particle separation $d$ (inset).
After stirring, the column integrated density $\tilde{n}_0(v)$ at the center of the cloud is reduced for $v>v_{c}$ compared to the unperturbed value, indicating heating.
For a superfluid gas (blue circles), the critical velocity $v_c$ can be determined from a bilinear fit (blue line) and  in a thermal cloud (red circles and line), no critical velocity can be observed. The data is acquired at {$B=806$\,G}, $a=13500\,a_0$ with $\tilde{n}_0=\unit{1.11}{\micro\metre^{-2}}$, $N=6100$ for the superfluid.}
\end{figure}

Here, we explore the stability of superfluids against external perturbation in the crossover from Bose-Einstein condensation (BEC) of composite bosons to Bardeen-Cooper-Schrieffer (BCS) pairing of fermions. An obstacle consisting of a small attractive potential is moved through an oblate superfluid gas. Above a critical velocity heating is observed, as shown in Fig. \ref{fig:scheme}.  For a pointlike weak perturbation, the Landau criterion $v_{c}=\min_{p}(\epsilon(p)/p)$ makes the direct connection between the critical velocity $v_{c}$ as a macroscopic observable and the microscopic excitations of the system with energy $\epsilon(p)$ and momentum $p$. One source of heating is the excitation of phonons. For these excitations, the Landau criterion predicts that the critical velocity equals the sound velocity $v_s$, which can be calculated within the Bogoliubov approximation for a weakly interacting Bose gas. Consequently, we measure $v_s$ as well by exciting and tracking density modulations. The obtained results are compared to the critical velocities.

Previously, $v_c$ has been measured in ultracold Bose and Fermi gases. Weakly interacting three-dimensional \cite{Ketterle1999} and two-dimensional \cite{Dalibard2012} BECs were probed with moving repulsive obstacle potentials and critical velocities of $10\,\%$ and $60\,\%$ of the Bogoliubov sound velocity were found. It is expected that vortex excitations limited $v_c$ \cite{Zwerger2000} since the healing length was much smaller than the obstacle size. In Fermi gases, $v_c$ was determined in the BEC-BCS crossover by subjecting the cloud to a moving optical lattice \cite{Ketterle2007}. A comparison with theory was performed at the universal point yielding $v_c\approx70\,\%\,v_s$.
The precise microscopic excitation mechanism is not fully understood yet, but theoretical analyses \cite{Stringari2009} suggested that it is quite different from the one relevant in our measurements.
In the crossover, $v_s$ was measured as well \cite{vs_exp}.
However, in those experiments no comparison to $v_c$ was made.

Due to the high optical resolution and low densities achieved in our apparatus, it is finally possible to manipulate and probe superfluids on their intrinsic length scales. The obstacle size is on the order of the healing length in the BEC regime, the coherence length in the BCS regime, and the inter-particle separation in the crossover. Our main results are shown in Fig. 2: they consist of measurements of $v_c$, $v_s$, and a detailed comparison with theory in the entire crossover. The results for $v_s$ are in very good agreement with the theoretical prediction. In the BEC regime, the critical velocity is found to be significantly smaller than $v_s$ but in excellent agreement with numerical simulations. The simulations take all experimental details into account and allow us to determine the origins of the reduction. Having validated the method in the BEC regime, our results in the strongly correlated regime may provide valuable benchmarks for theory. In the BCS regime, pair-breaking excitations are expected to limit $v_c$ and our results are in qualitative agreement. 

We prepare $^6$Li atoms with mass $m$ in a balanced mixture of the two lowest hyperfine states with a similar procedure as described in Ref. \cite{Moritz2011}. Ultimately, the atoms are trapped in a highly elliptical optical dipole trap  with a beam waist of $\unit{10}{\micro\metre}\times\unit{370}{\micro\metre}$ and a wavelength of $1064\, \nano \metre$. Typical trap frequencies are $\omega_{z}\approx2\pi\cdot\unit{550}{\hertz}$ and $\omega_{r}\approx2\pi\cdot\unit{30}{\hertz}$ in the vertical and radial direction. The radial confinement is mainly caused by the curvature of a radially symmetric magnetic field. We adjust the final evaporation to obtain a constant line of sight integrated central density of $\tilde{n}_0=(1.15 \pm 0.05)\, \micro\metre^{-2}$ per spin state. Depending on the interaction strength, this corresponds to a total atom number $N$ of $2500$ to $14000$ per spin state.
We estimate the systematic errors on atom numbers and densities to be approximately $\pm\,20\,\%$.
Although the vertical confinement dominates, effects caused by reduced dimensionality are negligible since $E_F/\hbar\omega_z>4.2$ in all measurements, where the Fermi energy $E_F$ and wavevector $k_F$ are defined as $E_{F}=\hbar^2k_F^2/2m=\hbar (\omega_r^2\omega_z\cdot6N)^{1/3}$. A measure for the temperature $T$ is provided by the observed condensate fractions in the BEC regime of approximately $80\,\%$. Since we observe no significant heating during magnetic field ramps, we use the theory in Ref. \cite{Castin2004} to estimate the temperature in the BCS regime, yielding values of $T/T_F \approx 7\,\%$. 

In the actual stirring experiment, a red-detuned laser beam forms an attractive potential. This obstacle traces out a circular trajectory with speed $v$ and radius $r=10\,\micro$m along lines of constant column density $\tilde{n}(r)\approx \tilde{n}_0$ within the superfluid core. The beam has a wavelength of $\unit{780}{\nano\metre}$ and is focused to a waist $w$ of $\unit{2.4}{\micro \metre} \times \unit{1.9}{\micro \metre}$, a size comparable to the interparticle distance $d=n^{-1/3}\approx\unit{1.5}{\micro\metre}$ at unitarity.
The relative column integrated density increase in the focus is approximately $85\,\%$.
The corresponding beam powers were adjusted depending on the interaction strength.

\begin{figure}
\includegraphics{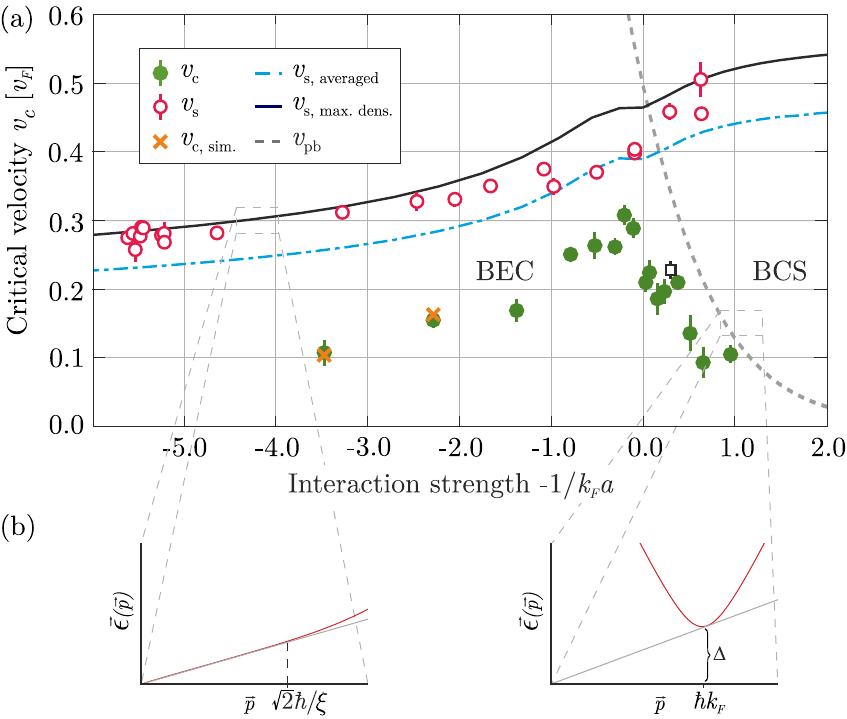}
\caption{\label{fig:results}
(a) Critical velocity $v_c$ (green filled circles) and speed of sound $v_s$ (red open circles) in units of the Fermi velocity $v_F$ throughout the BEC-BCS crossover. The error bars correspond to the fit errors. A statistical error for $v_c$ (black open square) was determined from five measurements. The simulated critical velocities are marked with crosses.
The solid (dot-dashed) curve is the theory prediction for $v_s$ assuming that the maximum (column averaged) density is relevant for sound propagation, see main text. The pair breaking velocity $v_{pb}$ providing an upper bound for $v_c$ in the BCS regime is plotted with a dashed line.
(b) Dispersion relations for the BEC and the BCS limiting cases (red) and the tangent to this curve from the origin to visualize the Landau criterion (grey).
}
\end{figure}

The stirring sequence proceeds as follows: first, the scattering length $a$ is set to the desired value by ramping the magnetic field to a value between {750\,G} and {890\,G} close to a broad Fesh\-bach resonance, followed by $\unit{50}{\milli\second}$ thermalization time. Next, the power of the moving obstacle beam is linearly ramped up within $\unit{10}{\milli\second}$ and the gas is stirred for $\unit{200}{\milli\second}$ before the power is linearly ramped down in $\unit{5}{\milli\second}$. After $\unit{100}{\milli\second}$ thermalization time the magnetic field is ramped to {680\,G} in $\unit{100}{\milli\second}$ and an in-situ absorption image of the atoms is acquired. 
We repeat this sequence typically ten times for each speed $v$  and extract the radially averaged and line of sight integrated density distribution $\tilde{n}(r)$ from the mean of those datasets, accounting for optical saturation effects \cite{Reinaudi2007}. 
Since the gas is well in the BEC regime at the time of imaging, we determine the central column density $\tilde{n}_0(v)$ as well as the condensate fraction  from a bimodal fit. Heating is indicated by a reduction in either, yet $\tilde{n}_0(v)$ is the more robust measure since evaporation upon heating can occur in our trap of finite depth.

We observe a significant reduction in $\tilde{n}_0(v)$ and hence heating only above a threshold velocity which we identify with the critical velocity as shown in Fig. \ref{fig:scheme}.
The exact value is obtained from a fit with a continuous bilinear function \cite{Ketterle2007}. It has a constant value of $\tilde{n}_0$ below $v_c$ and decreases linearly above, see blue line in Fig. \ref{fig:scheme}. The figure also shows that stirring within the thermal region of the cloud leads to heating for all obstacle speeds.

We determine the critical velocities for different interaction strengths $-1/k_{F}a$ throughout the whole BEC-BCS crossover and far into the BEC regime and plot them in units of the Fermi velocity $v_{F}$ in Fig. \ref{fig:results}(a). 
Qualitatively, the data shows a maximum of $v_c$ close to $1/k_{F}a=0$ and a decrease towards the BEC and the BCS side of the resonance, in agreement with Ref. \cite{Ketterle2007}. The absolute values range between $\unit{1.7}{\milli\metre\per\second}\leq v_c\leq\unit{6.3}{\milli\metre\per\second}$. For comparison we also measure the speed of sound $v_s$ by creating a small density excess in the center of the gas, releasing it and tracking the maximum of the outgoing circular density wave. Here, the stirrer beam is placed at the center of the gas, its power is adiabatically raised to values between $\unit{7}{\micro\watt}$ and $\unit{40}{\micro\watt}$ in $\unit{100}{\milli\second}$ and suddenly switched off.

To compare the experimental results with theoretical predictions, it is convenient to consider three regimes, the BEC, the strongly correlated regime, and the BCS regime.
In the latter ($-1/k_{F}a>1$), superfluids are formed from loosely bound Cooper pairs. 
The excitation spectrum is sketched in the \rhs of Fig. \ref{fig:results}(b). Pair breaking excitations limit the critical velocity to $m\, v^2_{pb}={\left(\Delta^{2}+\mu^{2}\right)^{1/2}-\mu}$ \cite{Stringari2006}. 
The pair breaking velocity $v_{pb}$ is plotted as the dashed line in Fig. \ref{fig:results}(a), where we determined the gap $\Delta$ and the chemical potential $\mu$ at $T=0$ by solving the mean field gap the number equations numerically \cite{Strinati1998,Stringari2008}. The curve can be extended into the strongly correlated regime, where no simple theoretical description exists. Here, the mean field approach can at least provide a rough estimate for $v_c$ and our data appears to be in qualitative agreement. We expect temperature effects to be small since $T/T_c<0.5$ \cite{Muehlschlegel1959}.

Before discussing the strongly correlated regime in depth, which is theoretically largely inaccessible and hence particularly interesting,  we benchmark our experiment against theory. In the BEC regime ($-1/k_{F}a<-1$), the gas forms a molecular BEC of tightly bound dimers. Within Bogoliubov theory the dispersion relation is linear at low momenta with a slope $v_s$, see \lhs of Fig. \ref{fig:results}(b), and $v_c$ should equal $v_s$. The measured sound velocities are in very good agreement with the two theoretical predictions shown in Fig. \ref{fig:results}(a). When the sound wavelength is large compared to the vertical extent of the cloud, the wave effectively probes the column averaged density (dot-dashed line), provided the gas is fully hydrodynamic \cite{vs_theo}.
Otherwise, the wavefront observed should be the one travelling with the speed determined by the maximum density along the z-direction (solid line). Since the gas is only partially hydrodynamic in the vertical direction, we expect the experimental data to lie between the two curves. We note that the measurements of $v_s$ presented here probe a new regime since all previous experiments determining $v_s$ were performed in prolate gas clouds \cite{vs_exp} described by effectively one-dimensional hydrodynamics \cite{vs_theo}. The theory curves for $v_s$ are obtained by taking thermodynamic derivatives \cite{Salasnich2005} of the equation of state calculated in numerically exact zero-temperature quantum Monte Carlo simulations \cite{Giorgini2004}. The homogeneous theory is applied using the local density approximation: the density distribution in the trap, given by the equation of state, is used to calculate $k_F$ and $v_F$ of the corresponding trapped clouds \cite{k_F}. 
Temperature effects should be small since the temperatures in the experiment are smaller than the mean field energy in the BEC regime and the Fermi temperature in the BCS regime \cite{Heiselberg2006}.

\begin{figure}
	\includegraphics{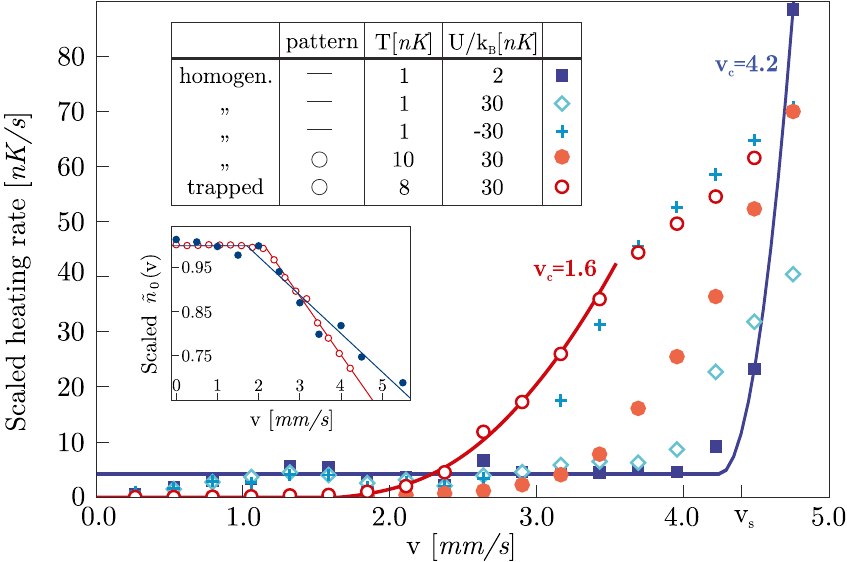}
	\caption{\label{fig:simulation}
		The simulated heating rates normalized by the stirrer depth $U^2$. The complexity is gradually increased: blue squares depict the idealized case of a very cold homogeneous sample stirred with linear pattern. The relative density excess $\eta$ in the weak stirrer potential $U=k_B\cdot\unit{2}{\nano\kelvin}$ is only $3\,\%$.
		For all datasets, the Bogoliubov result for $v_s$ is $\unit{4.4}{\milli\metre\per\second}$.
		The red open circles depict a simulation of the experimental case: a trapped sample is stirred circularly with a stirrer of realistic depth. A lower temperature is chosen for technical reasons. 
		Here, the y-axis scaling factor is one.
		In the inset, the results for the heating observed in the central column density $\tilde{n}_0(v)$ are compared. We find very good agreement between the experimental (blue filled circles) and the simulated results (red open circles). The bilinear fits to extract $v_c$ are shown with solid lines. In the inset $U=k_B\cdot\unit{35}{\nano\kelvin}$. 
	}
\end{figure}

In order to understand the critical velocity in the BEC regime, we perform simulations and identify the factors reducing $v_c$. These are the finite temperature, the inhomogeneous density profile along the strongly confined direction, the circular instead of linear motion of the stirrer, and to a lesser degree the finite depth of the obstacle potential. We use a classical field method, which is the limiting case of the truncated Wigner method used in Ref. \cite{AmyMathey2014}. The time evolution of an ensemble of complex-valued fields is calculated using classical equations of motion. The initial states are generated from a grand canonical ensemble via a classical Metropolis algorithm.  We employ a real-space representation on a lattice with $60\times60\times3$  ($140\times140\times11$) sites for the simulation of homogeneous (trapped) systems. The discretization length is $\unit{1}{\micro\metre}$. All simulations are performed with the same stirring time, stirrer beam size, dimer-dimer scattering length $a_{DD}=0.6\times 3634\,a_0$, and density $n_{3D}=\unit{0.486}{\micro\metre^{-3}}$ (and column density in the trapped case) as the experimental data point at $-1/k_Fa\approx-3.5$. 
When choosing all remaining parameters, \ie temperature, confining potential, stirrer depth, and motion in accordance with the experiment, we reproduce the experimentally measured  $v_c$. 
To disentangle the various features of the system that influence these measurements, it is instructive to start with an idealized case: a  homogeneous gas at a low temperature of $\unit{1}{\nano\kelvin}$, stirred along a linear path. In this case, the heating rate increases steeply at a critical velocity which is approximately $v_s$ as shown in Fig. \ref{fig:simulation}. To determine $v_c$, the fit function $A\cdot\left(v^2-v_c^2\right)^2/v+B$ is used for $v>v_c$ \cite{Ptaevskii2004}, with the free parameters $A$, $B$ and $v_c$.
The simulated heating rates are in good agreement with the second order perturbation theory that predicts a scaling with $U^2$. Moreover, by increasing the stirrer depth $U$, we observe that the extracted $v_c$ is slightly reduced.
These results demonstrate that we work with relatively weak perturbations and that vortex excitations do not limit $v_c$ \cite{Zwerger2000}, in contrast to previous experiments in 3D \cite{Ketterle1999} and 2D atomic BECs \cite{Dalibard2012}. The simulations also show that attractive stirrer potentials are preferable to realize a stirrer. For larger repulsive potentials \cite{Ketterle1999,Dalibard2012} the inherent density reduction strongly reduces the observed critical velocity as shown in Fig. \ref{fig:simulation}.

Next, the effects of the finite experimental temperature and of the circular motion of the stirrer are investigated. The simulations show that both features reduce $v_c$ by approximately $15\,\%$. Having both present simultaneously causes a small further reduction of $v_c$. The reduction at finite temperature might be due to vortex-antivortex excitations, or rotonic precursors of them. As the temperature is increased above the mean field energy, density fluctuations increase and vortices can nucleate at points of minimal density. That the circular motion can reduce $v_c$ can be seen in perturbation theory performed in momentum space: here, the motion of the perturbation consists of a distribution of velocities rather than a single velocity.

Finally, we perform a simulation of an inhomogeneous system in a trap, with a realistic temperature and a circular stirring motion. 
The simulated critical velocity of $1.6(1)$ mm/s agrees excellently with the experimentally measured value of $1.7(3)$ mm/s, see Fig. \ref{fig:results}. 
We believe that the additional reduction of $39\,\%$ with respect to the homogeneous simulation result is mainly due to probing lower density regions along the stirrer axis.
The results for the central column densities are in good agreement as well, see inset of Fig. \ref{fig:simulation}, considering the experimental signal to noise.

We now turn to the strongly correlated regime. Due to the lack of a small parameter, perturbation theories are inaccurate and the quasiparticle description breaks down. Hence, the velocities $v_s$ and $v_{pb}$ associated with phonon creation and Cooper pair breaking excitations can only provide upper limits for $v_c$. We are not aware of a prediction for $v_c$, even at the universal point where $|a|\rightarrow \infty$. 
The largest value for $v_c$ we observe is $v_{c}=0.31(2)\,v_F$, close to the universal point, see Fig. \ref{fig:results}. Reference \cite{Ketterle2007} found a value of $v_c=0.31 \,v_F$ using a different excitation mechanism. These values are considerably smaller than the corresponding $v_s\approx0.40(1)\,v_F$ we measure and the theory prediction $v_s=\xi_B^{1/4}/\sqrt{3}\,v_F=0.45\,v_F$ \cite{Stringari2008,Bertsch} employing the local density approximation. 
Very recently, a critical velocity of $v_c=0.42^{+0.05}_{-0.11}\,v_F$ was observed in an elongated $^6$Li gas oscillating with respect to a $^7$Li BEC \cite{Salomon2014}. Here,  the onset of heating is predicted to occur for a relative velocity that equals the sum of the individual sound velocities \cite{Castin2014}.

In conclusion, we have demonstrated the breakdown of superfluidity due to moving obstacle across the BEC-BCS transition, for the first time in close analogy to Landau's Gedankenexperiment. We compare the results with theoretical predictions throughout and achieve quantitative understanding in the BEC regime by performing numerical simulations. Pointlike defects also play a role in strongly correlated high temperature superconductors. The experiment presented here provides the opportunity to isolate relevant effects in a very clean and controllable environment. Of particular interest for future studies are strongly correlated two-dimensional superfluids. 

We thank N. Strohmaier, J. H. Drewes, and F. Wittk\"otter for their contributions to early stages of this experiment and J. Dalibard, C. Weitenberg, W. Zwerger, and especially L. Tarruell for stimulating discussions. This work has been supported financially by the Deutsche Forschungsgemeinschaft within SFB 925, GRK 1355, the Hamburg Centre for Ultrafast Imaging, and by the Landesexzellenzinitiative Hamburg, which is supported by the Joachim Herz Stiftung.



\begin{thebibliography}{1}
\bibitem{k_F} Please note that the commonly accepted definition of $k_F$ used also here is not trivially connected to the density.
\end{thebibliography}
\end{document}